\documentstyle[12pt,epsfig]{article}
\def\thebibliography#1{\leftline{\large \it References}\list
  {[\arabic{enumi}]}{\settowidth\labelwidth{[#1]}\leftmargin\labelwidth
    \advance\leftmargin\labelsep
    \usecounter{enumi}}
    \def\newblock{\hskip .11em plus .33em minus .07em}
    \sloppy\clubpenalty4000\widowpenalty4000}

\newcommand{\be}{\begin{eqnarray}}
\newcommand{\ba}{\begin{array}}
\newcommand{\ea}{\end{array}}
\newcommand{\ee}{\end{eqnarray}}
\newcommand{\dslash}{\partial \hskip -0.5em /}

\newcommand{\Tr}{{\rm Tr}}

\newcommand{\La}{{\cal L}}
\newcommand{\A}{{\cal A}}

\newcommand{\bjlim}{{\stackrel{\scriptstyle{\rm Bj}}
{\textstyle\longrightarrow}}}

\begin{document}
\rightline{February 1997}
\vskip 1.0truecm
\centerline{\Large\bf Nucleon Structure Functions from a}
\vskip 0.5truecm
\centerline{\Large\bf Chiral Soliton$^{*}$}
\baselineskip=18 true pt
\vskip 1.0cm
\centerline{H.\ Weigel$^{a)}$, L.\ Gamberg$^{a,b)}$, 
and H.\ Reinhardt$^{a,c)}$}
\vskip 0.5cm
\centerline{$^{a)}$Institute for Theoretical Physics}
\centerline{T\"ubingen University, Auf der Morgenstelle 14}
\centerline{D-72076 T\"ubingen, Germany}
\vskip 0.4cm
\centerline{$^{b)}$Department of Physics and Astronomy}
\centerline{University of Oklahoma, 440 West Brooks}
\centerline{Norman, Ok. 73019, USA}
\vskip 0.4cm
\centerline{$^{c)}$Center for Theoretical Physics} 
\centerline{Laboratory for Nuclear Science and Department of Physics}
\centerline{Massachusetts Institute of Technology} 
\centerline{Cambridge, Ma. 02139, USA}
\vskip 1.5cm
\baselineskip=16pt
\centerline{\bf ABSTRACT}
\vskip 0.5cm
Nucleon structure functions are studied  within the chiral soliton 
approach to the boson\-ized Nambu--Jona--Lasinio model. The valence 
quark approximation is employed which is justified for moderate 
constituent quark masses ($\sim 400{\rm MeV}$) as the contribution 
of the valence quark level dominates the predictions of nucleon 
properties. As examples the unpolarized structure functions for 
the ${\nu}p$ and ${\bar \nu}p$ scattering and the structure 
functions entering the Gottfried sum rule are discussed. For the 
latter the model prediction is found to reasonably well agree 
with a corresponding low--scale parametrization of the empirical 
data.
\vskip 0.5cm
\leftline{\it PACS: 12.39.Fe, 12.39.Ki.}
\vskip 0.5cm
\leftline{\it Keywords: \parbox[t]{15cm}{Nucleon Structure Functions,
Lepton--Nucleon Scattering,\\
Nambu--Jona--Lasinio Model, Chiral Soliton.}}

\vfill
\noindent
$^{*}$
{\footnotesize{Supported in part by the Deutsche Forschungsgemeinschaft
(DFG) under contract Re 856/2-2 and by funds provided by the 
U.S. Department of Energy (D.O.E.) under cooperative agreement 
\# DE--FC02--94ER40818.}}\\
\eject

\normalsize
\baselineskip16pt
\leftline{\large\it 1. Introduction}
\smallskip

It has been a long standing puzzle how to establish the connection 
between the chiral soliton picture of the baryon, which essentially 
views baryons as mesonic lumps, and the quark parton model, which 
regards baryons as composites of almost non--interacting, point--like 
quarks. While the former has been quite successful in describing 
static properties of the nucleon, the latter, being firmly 
established within the context of deep inelastic scattering (DIS), has 
been quite successful in predicting the spin average DIS nucleon 
structure functions. The apparent difference between models for the 
nucleon like the bag model \cite{Ch74}, which have previously been 
employed to study structure functions \cite{Ja75}--\cite{So94}, 
and soliton  models is the fact that in the latter the nucleon 
wave--function only appears as a collectively excited (topologically) 
non--trivial meson configuration rather than as a product of Dirac 
spinors. In this letter we calculate structure functions in the 
Nambu--Jona--Lasinio (NJL) \cite{Na61} chiral soliton model 
\cite{Re89,Al96} where the hadronic currents are formally described 
in terms of quark degrees of freedom which themselves are functionals 
of the solitonic meson fields. Since the present study is the first 
step towards computing nucleon structure functions from a chiral 
soliton we will adopt a simplifying valence quark type of approximation 
(to be defined after eq (\ref{gendef1})) and leave a more complete 
exploration to future studies. 

As in the original study \cite{Ja75} of structure functions for 
localized field configurations, the structure functions are most
easily accessible when the current operator is at most quadratic in 
the fundamental fields and the propagation of the interpolating 
field can be regarded as free. Although the latter approximation is 
well justified in the Bjorken limit the former condition is 
difficult to satisfy in soliton models built from mesonic 
fields. In such models the soliton is a non--perturbative object 
involving all orders of the fundamental pion field. Hence the current 
operator is not confined to quadratic order. In models where mesons 
are fundamental fields ({\it e.g.} the Skyrme model \cite{Sk61,Ad83}, 
the chiral quark model of ref. \cite{Bi85} or the chiral bag model 
\cite{Br79}\footnote{In the cloudy bag model the contribution of the 
pions to structure functions has been treated perturbatively 
\cite{Sa88,Sc92}.}) structure functions are exceedingly difficult to 
obtain. In this respect the chirally invariant NJL model 
is advantageous because it is entirely defined in terms of quark 
degrees of freedom. This makes the evaluation of the required 
commutator (see eq (\ref{deften}) below) feasible. Nevertheless 
the quark currents become uniquely (up to regularization) defined 
functionals of the meson fields. The Lagrangian of the 
NJL model reads
\be
\La = \bar q (i\dslash -  m^0 ) q +
      2G_{\rm NJL} \sum _{i=0}^{3}
\left( (\bar q \frac {\tau^i}{2} q )^2
      +(\bar q \frac {\tau^i}{2} i\gamma _5 q )^2 \right) .
\label{NJL}
\ee
Here $q$, $\hat m^0$ and $G_{\rm NJL}$ denote the quark field, the 
current quark mass and a dimensionful coupling constant, respectively.
Functional bosonization \cite{Eb86} yields the action
\be
\A&=&\Tr_\Lambda\log(iD)+\frac{1}{4G_{\rm NJL}}
\int d^4x\ {\rm tr}
\left(m^0\left(M+M^{\dag}\right)-MM^{\dag}\right)\ , 
\label{bosact} \\
D&=&i\dslash-\left(M+M^{\dag}\right)
-\gamma_5\left(M+M^{\dag}\right)\ .
\label{dirac}
\ee
The composite scalar ($S$) and pseudoscalar ($P$) meson fields 
are contained in $M=S+iP$, and appear as quark--antiquark bound 
states. For regularization, which is indicated by the cut--off 
$\Lambda$, we will adopt the proper--time scheme \cite{Sch51}.
The free parameters of the model are the current quark mass $m^0$, 
the coupling constant $G_{\rm NJL}$ and the cut--off $\Lambda$.
When expanding $\A$ to quadratic order in $M$ these parameters are 
related to the pion mass, $m_\pi=135{\rm MeV}$ and decay constant,
$f_\pi=93{\rm MeV}$. This leaves one undetermined parameter which we 
choose to be the vacuum expectation value $m=\langle M\rangle$. For 
apparent reasons $m$ is called the constituent quark mass. It is 
related to $m^0$, $G_{\rm NJL}$ and $\Lambda$ via the gap--equation, 
{\it i.e.} the equation of motion for the scalar field $S$. The 
occurrence of this vacuum expectation value reflects the spontaneous 
breaking of chiral symmetry.

We will approach the computation of structure functions in the 
NJL model by first briefly reviewing the kinematics of the Bjorken limit 
and the NJL soliton in sections 2 and 3, respectively. In section
4 we will work out the valence quark approximation to the unpolarized 
structure functions. The numerical results will be presented in 
section 5. Finally section 6 not only serves to summarize these 
studies but also to propose further explorations.

\bigskip
\leftline{\large\it 2. Kinematics}
\smallskip

The starting point for computing nucleon structure functions
is the hadronic tensor
\be
W^{ab}_{\mu\nu}(q)=\frac{1}{4\pi}\int d^4 \xi \ 
{\rm e}^{iq\cdot\xi}
\langle N |\left[J^a_\mu(\xi),J^{b{\dag}}_\nu(0)\right]|N\rangle \ .
\label{deften}
\ee
Here $|N\rangle$ refers to the nucleon state and 
$J^a_\mu(\xi)={\bar q}(\xi)\gamma_\mu t_a q(\xi)$ is the hadronic 
vector current. In the context of weak interactions we take
$J^a_\mu(\xi)={\bar q}(\xi)\gamma_\mu(1-\gamma_5)t_a q(\xi)$.
We denote by $t_a\ (a=0,..,3)$ the flavor operators, which 
have to be chosen appropriately for the process under 
consideration\footnote{For simplicity we omit the flavor index when 
not relevant.}. The unpolarized structure functions are related to 
the symmetric piece, $W^{(S)}_{\mu\nu}=(W_{\mu\nu}+W_{\nu\mu})/2$,
which is parametrized by two scalar form factors, 
\be
W^{(S)}_{\mu\nu}(q)=
\left(-g_{\mu\nu}+\frac{q_\mu q_\nu}{q^2}\right)
W_1(x_{\rm Bj},Q^2)+
\left(P_\mu-q_\mu\frac{P\cdot q}{q^2}\right)
\left(P_\nu-q_\nu\frac{P\cdot q}{q^2}\right)
W_2(x_{\rm Bj},Q^2)\ .
\label{defff}
\ee
Here $P_\mu$ refers to the nucleon momentum and $Q^2=-q^2$. 
Furthermore quantities suitable to study the Bjorken scaling have 
been introduced: $\nu={P\cdot q}/{M_N}$ and
$x_{\rm Bj}={Q^2}/{2M_N\nu}$.
Introducing the projection operators \cite{So94}
\be
\Lambda^{\mu\nu}_1=\frac{1}{2}
\left[-g^{\mu\nu}+\frac{\eta}{M_N^2}P^\mu P^\nu\right]\ , \quad
\Lambda^{\mu\nu}_2=\frac{1}{2}
\left[-g^{\mu\nu}+\frac{3\eta}{M_N^2}P^\mu P^\nu\right]
\label{proj1}
\ee
with $\eta=2M_N\nu x_{\rm Bj}/(2M_N\nu x_{\rm Bj}+\nu)$
enables one to straightforwardly extract the form factors
\be
W_i(x_{\rm Bj},Q^2)=\Lambda^{\mu\nu}_i W^{(S)}_{\mu\nu}(q)
\ , \quad i=1,2\ .
\label{proj2}
\ee
When discussing Bjorken scaling a slightly different definition 
of the form factors 
\be
F_1(x_{\rm Bj},Q^2)=M_N W_1(x_{\rm Bj},Q^2) 
\quad {\rm and}\quad
F_2(x_{\rm Bj},Q^2)=\nu W_2(x_{\rm Bj},Q^2)
\label{deff12}
\ee
is commonly considered. The Bjorken limit corresponds to the 
kinematical regime 
\be
q_0=|\mbox{\boldmath $q$}| - M_N x_{\rm Bj}
\quad {\rm with}\quad
|\mbox{\boldmath $q$}|\rightarrow \infty \ .
\label{bjlimit}
\ee
Finally the structure functions are obtained as the Bjorken limit
of the form factors (\ref{deff12})
\be
F_i(x_{\rm Bj})=\lim_{\rm Bj} F_i(x_{\rm Bj},Q^2)\ , \qquad i=1,2\ .
\label{defstfct}
\ee

\bigskip
\leftline{\large\it 3. The Nucleon State in the NJL Model}
\smallskip

As the NJL model soliton has exhaustively been discussed in 
a recent review article \cite{Al96} we only present those features, 
which are relevant for the computation of the structure functions.

The chiral soliton is given by the hedgehog configuration of the 
meson fields
\be
M_{\rm H}(\mbox{\boldmath $x$})=m\ {\rm exp}
\left(i\mbox{\boldmath $\tau$}\cdot{\hat{\mbox{\boldmath $x$}}}
\Theta(r)\right)\ .
\label{hedgehog}
\ee
In order to compute the functional trace in eq (\ref{bosact}) for this 
static configuration a Hamilton operator, $h$ is extracted from the 
Dirac operator (\ref{dirac}). That is, $D=i\gamma_0(\partial_t-h)$ 
with
\be
h=\mbox{\boldmath $\alpha$}\cdot\mbox{\boldmath $p$}+m\ 
{\rm exp}\left(i\gamma_5\mbox{\boldmath $\tau$}
\cdot{\hat{\mbox{\boldmath $x$}}}\Theta(r)\right)\ .
\label{hamil}
\ee
We denote the eigenvalues and eigenfunctions of $h$ by
$\epsilon_\mu$ and $\Psi_\mu$, respectively. In the proper time 
regularization scheme the NJL model energy functional is found to 
be \cite{Re89,Al96}
\be
E[\Theta]=
\frac{N_C}{2}\epsilon_{\rm v}
\left(1+{\rm sgn}(\epsilon_{\rm v})\right)
+\frac{N_C}{2}\int^\infty_{1/\Lambda^2}
\frac{ds}{\sqrt{4\pi s^3}}\sum_\nu{\rm exp}
\left(-s\epsilon_\nu^2\right)+
m_\pi^2 f_\pi^2\int d^3r  \left(1-\Theta(r)\right) ,
\label{efunct}
\ee
with $N_C=3$ being the number of color degrees of freedom. 
The subscript ``${\rm v}$" denotes the valence quark level. This state 
is the distinct level bound in the soliton background with
$-m<\epsilon_{\rm v}<m$. The chiral angle, $\Theta(r)$, is
obtained by self--consistently extremizing $E[\Theta]$ \cite{Re88}.

Nucleon states possessing good spin and isospin quantum numbers are 
generated from the soliton by taking the zero--modes to be time 
dependent \cite{Ad83}
\be
M(\mbox{\boldmath $x$},t)=
A(t)M_{\rm H}(\mbox{\boldmath $x$})A^{\dag}(t)\ ,
\label{collrot}
\ee
which introduces the collective coordinates $A(t)\in SU(2)$. The 
action functional is expanded \cite{Re89} in the angular velocities 
\be 
2A^{\dag}(t)\dot A(t)=
i\mbox{\boldmath $\tau$}\cdot\mbox{\boldmath $\Omega$}.
\label{angvel}
\ee
In particular the valence quark wave--function receives a first 
order perturbation
\be
\Psi_{\rm v}(\mbox{\boldmath $x$},t)=
{\rm e}^{-i\epsilon_{\rm v}t}A(t)
\left\{\Psi_{\rm v}(\mbox{\boldmath $x$})
+\frac{1}{2}\sum_{\mu\ne{\rm v}}
\Psi_\mu(\mbox{\boldmath $x$})
\frac{\langle \mu |\mbox{\boldmath $\tau$}\cdot
\mbox{\boldmath $\Omega$}|{\rm v}\rangle}
{\epsilon_{\rm v}-\epsilon_\mu}\right\}=:
{\rm e}^{-i\epsilon_{\rm v}t}A(t)
\psi_{\rm v}(\mbox{\boldmath $x$}).
\label{valrot}
\ee
Here $\psi_{\rm v}(\mbox{\boldmath $x$})$ refers to the spatial part 
of the body--fixed valence quark wave--function with the rotational 
corrections included. Nucleon states $|N\rangle$ are obtained 
by canonically quantizing the collective coordinates, $A(t)$. By 
construction these states live in the Hilbert space of a rigid rotator. 
The eigenfunctions are Wigner $D$--functions
\be
\langle A|N\rangle=\frac{1}{2\pi}
D^{1/2}_{I_3,-J_3}(A)\ ,
\label{nwfct}
\ee
with $I_3$ and $J_3$ being respectively the isospin and spin 
projection quantum numbers of the nucleon. 

\bigskip
\leftline{\large\it 4. Unpolarized Structure Functions in the 
Valence Quark Approximation}
\smallskip

The starting point for the computation of the unpolarized structure
functions is the hadronic tensor in the form suitable for localized
fields \cite{Ja75}
\be
W^{lm}_{\mu\nu}(q)&=&\zeta\int \frac{d^4k}{(2\pi)^4} \
S_{\mu\rho\nu\sigma}\ k^\rho\
{\rm sgn}\left(k_0\right) \ \delta\left(k^2\right)
\int_{-\infty}^{+\infty} dt \ {\rm e}^{i(k_0+q_0)t}
\nonumber \\ && \hspace{1cm}
\times \int d^3x_1 \int d^3x_2 \
{\rm exp}\left[-i(\mbox{\boldmath $k$}+\mbox{\boldmath $q$})\cdot
(\mbox{\boldmath $x$}_1-\mbox{\boldmath $x$}_2)\right]
\nonumber \\ && \hspace{1cm}
\times \langle N |\left\{
{\bar \Psi}(\mbox{\boldmath $x$}_1,t)t_l t_m\gamma^\sigma
\Psi(\mbox{\boldmath $x$}_2,0)-
{\bar \Psi}(\mbox{\boldmath $x$}_2,0)t_m t_l\gamma^\sigma
\Psi(\mbox{\boldmath $x$}_1,t)\right\}| N \rangle .
\label{stpnt}
\ee
Here $S_{\mu\rho\nu\sigma}=g_{\mu\rho}g_{\nu\sigma}
+g_{\mu\sigma}g_{\nu\rho}-g_{\mu\nu}g_{\rho\sigma}$ denotes the 
symmetric combination of $\gamma_\mu\gamma_\rho\gamma_\nu$ and
$\zeta=1(2)$ for the structure functions associated with the 
vector (weak) current. As explained in the preceding section the 
matrix element between the nucleon states ($|N\rangle$) is to be 
taken in the space of the collective coordinates. In deriving the 
expression (\ref{stpnt}) the {\it free} correlation function for 
the intermediate quark fields has been assumed \cite{Ja75}. This 
reduces the commutator $[J_\mu(\mbox{\boldmath $x$}_1,t),
J^{\dag}_\nu(\mbox{\boldmath $x$}_2,0)]$ 
of the quark currents in the definition (\ref{deften}) to objects 
which are merely bilinear in the quark fields. In the Bjorken limit 
(\ref{bjlimit}) the momentum, $k$, of the intermediate quark state 
is highly off--shell and hence not sensitive to momenta typical for 
the soliton configuration. Thus the use of the free correlation 
function is a good approximation in this kinematical regime.
Accordingly, the intermediate quark states are taken to be massless,
{\it cf.} eq (\ref{stpnt}). In the next step the form factors $W_i$ 
are extracted according to eq (\ref{proj2}). Noting that 
$\Lambda_1^{\mu\nu}S_{\mu\rho\nu\sigma}\bjlim g_{\rho\sigma}\ $
and 
$\Lambda_2^{\mu\nu}S_{\mu\rho\nu\sigma}\bjlim \eta g_{\rho\sigma}\ $
the Callan--Gross relation follows immediately, {\it i.e.} 
$F_2(x_{\rm Bj})=2x_{\rm Bj}F_1(x_{\rm Bj})$. It thus suffices 
to only consider the structure function $F_1(x_{\rm Bj})$.

Since the NJL model is formulated in terms of quark degrees of 
freedom, quark bilinears as in eq (\ref{stpnt}) can be computed from 
the functional
\be
\hspace{-1cm}
\langle {\bar q}(x)\Gamma q(y) \rangle
&=&\int D{\bar q} Dq \ {\bar q}(x)\Gamma q(y)\ 
{\rm exp}\left(i \int d^4x^\prime\ \La\right)
\nonumber \\ &&\hspace{-3cm}
=\frac{\delta}{i\delta\alpha(x,y)}\int D{\bar q} Dq \
{\rm exp}\left(i\int d^4x^\prime d^4y^\prime
\left[ \delta^4(x-y)\La+
\alpha(x^\prime,y^\prime){\bar q}(x^\prime)\Gamma
q(y^\prime)\right]\right)\Big|_{\alpha(x,y)=0}\ ,
\label{gendef}
\ee
where $\Gamma$ is a suitable Dirac and/or isospin matrix. The 
introduction of the bilocal source $\alpha(x,y)$ facilitates 
the  functional bosonization upon which eq (\ref{gendef}) 
takes the form
\be
\frac{\delta}{\delta\alpha(x,y)}{\rm Tr}_\Lambda{\rm log}
\left(\delta^4(x-y)D+\alpha(x,y)\Gamma)\right)
\Big|_{\alpha(x,y)=0}\ .
\label{gendef1}
\ee
The operator $D$ is defined in eq (\ref{dirac}). From this 
discussion it is obvious that structure functions are most easily 
obtained within models which can completely be formulated in terms 
of quark degrees of freedom where the form of the current operator 
is not altered by the interactions\footnote{Otherwise matrix elements 
of operators have to be computed, which are more complicated than 
the bilocal quark bilinear ${\bar q}(x)\Gamma q(y)$.}.

The correlation $\langle {\bar q}(x)\Gamma q(y) \rangle$ depends on 
the angle between $\mbox{\boldmath $x$}$ and $\mbox{\boldmath $y$}$.
Since in general the functional (\ref{gendef}) involves quark states 
of all angular momenta ($l$) a technical difficulty arises because 
the angular dependence has to be treated numerically. The major 
purpose of the present letter is to demonstrate that structure 
functions can be computed from a chiral soliton. With this in mind we 
will adopt the valence quark approximation where only quark orbital 
angular momenta up to $l=2$ are relevant. From a physical point of 
view this approximation is justified at least for small constituent 
quark masses because in that parameter region the nucleon properties 
are dominated by the valence quark contribution \cite{Al96}. We 
define the valence quark approximation to the structure functions 
by restricting the quark configurations in (\ref{gendef}) to the 
iso--rotating valence quark wave--function (\ref{valrot}), accordingly 
substituting the valence quark wave--function (\ref{valrot}) into 
eq (\ref{stpnt}). 

When extracting the structure function $F_1(x_{\rm Bj})$ the integrals 
over the time coordinate can readily be done yielding the conservation 
of energy for forward and backward moving intermediate quarks. Carrying 
out the integrals over $k_0$ and $k=|\mbox{\boldmath $k$}|$ yields for 
the isovector part of the structure function
\be
\hspace{-1cm}
F^{I=1}_1(x_{\rm Bj})&=&\zeta N_C\frac{M_N}{2\pi}\langle N |
D_{3i} \int d\Omega_{\mbox{\boldmath $k$}} k^2
\Bigg\{\tilde\psi_{\rm v}^{\dag}(\mbox{\boldmath $p$})
\left(1+\mbox{\boldmath $\alpha$}\cdot
{\hat{\mbox{\boldmath $k$}}}\right)
\tau_i\tilde\psi_{\rm v}(\mbox{\boldmath $p$})
\Big|_{k=q_0+\epsilon_{\rm v}}
{\rm tr}\left(\tau_3 t_l t_m\right)
\nonumber \\ && \hspace{3cm}
-\tilde\psi_{\rm v}^{\dag}(-\mbox{\boldmath $p$})
\left(1+\mbox{\boldmath $\alpha$}\cdot
{\hat{\mbox{\boldmath $k$}}}\right)
\tau_i\tilde\psi_{\rm v}(-\mbox{\boldmath $p$})
\Big|_{k=q_0-\epsilon_{\rm v}}
{\rm tr}\left(\tau_3 t_m t_l\right)
\Bigg\} |N\rangle\ ,
\label{valapp}
\ee
where $\mbox{\boldmath $p$}=\mbox{\boldmath $k$}+
\mbox{\boldmath $q$}$. $N_C$ appears as a multiplicative factor 
because the functional trace (\ref{gendef1}) includes the color 
trace as well. Furthermore the Fourier transform of the 
valence quark wave--function
\be
\tilde\psi_{\rm v}(\mbox{\boldmath $p$})=\int \frac{d^3x}{4\pi}\
\psi_{\rm v}(\mbox{\boldmath $x$})\
{\rm exp}\left(i\mbox{\boldmath $p$}\cdot
\mbox{\boldmath $x$}\right)
\label{ftval}
\ee
has been introduced. The isoscalar part of the structure function, 
$F^{I=0}_1(x_{\rm Bj})$, is straightforwardly obtained from eq 
(\ref{valapp}) by replacing $D_{3i}$ with unity and omitting the 
isospin matrices $\tau_i$. Note that the wave--function $\psi_{\rm v}$
contains an implicit dependence on the collective coordinates through
the angular velocity $\mbox{\boldmath $\Omega$}$, {\it cf.} eq 
(\ref{valrot}). 

The dependence of the wave--function 
$\tilde\psi(\pm\mbox{\boldmath $p$})$ on the integration variable 
${\hat{\mbox{\boldmath $k$}}}$ is only implicit. In order to carry out 
this integration it is most convenient to choose the 
external momentum along the $z$--axis, {\it i.e.}
$\mbox{\boldmath $q$}=q{\hat{\mbox{\boldmath $z$}}}$. In the Bjorken 
limit the integration variables may then be changed to \cite{Ja75}
\be
k^2 \ d\Omega_{\mbox{\boldmath $k$}} =
p dp\ d\Phi\ , \qquad p=|\mbox{\boldmath $p$}|\ ,
\label{intdp}
\ee
where $\Phi$ denotes the azimuth--angle between
$\mbox{\boldmath $q$}$ and $\mbox{\boldmath $p$}$. 
The lower bound for the $p$--integral is adopted when 
$\mbox{\boldmath $k$}$ and $\mbox{\boldmath $q$}$ are anti--parallel: 
$p^{\rm min}_\pm=|M_N x_{\rm Bj}\mp \epsilon_{\rm v}|$ for 
$k=q_0\pm\epsilon_{\rm v}$, respectively. The wave--function
$\tilde\psi(\pm\mbox{\boldmath $p$})$ acquires its dominant 
support for $p\le M_N$. Hence the integrand is different from 
zero only when $\mbox{\boldmath $q$}$ and $\mbox{\boldmath $k$}$
are anti--parallel and we may take 
${\hat{\mbox{\boldmath $k$}}}=-{\hat{\mbox{\boldmath $z$}}}$.
This is nothing but the light--cone description for computing 
the structure functions \cite{Ja91}.
The valence quark state possesses positive parity yielding
$\tilde\psi(-\mbox{\boldmath $p$})=\gamma_0
\tilde\psi(\mbox{\boldmath $p$})$. We now have arrived at the 
final expression for the isoscalar and isovector parts of the 
unpolarized structure function in the valence quark approximation
\be
F^{I}_1(x_{\rm Bj})&=&
\zeta\left(F^{I}_{+}(x_{\rm Bj})-F^{I}_{-}(x_{\rm Bj})\right)
\nonumber \\
F^{I=0}_{\pm}(x_{\rm Bj})&=&N_C\frac{M_N}{2\pi}\langle N|
\int^\infty_{M_N|x_\mp|}p dp \int_0^{2\pi}d\Phi\
\tilde\psi_{\rm v}^{\dag}(\mbox{\boldmath $p$}_\mp)
\left(1\mp\alpha_3\right)\tilde\psi_{\rm v}(\mbox{\boldmath $p$}_\mp)
|N\rangle {\rm tr}[t_l t_m]
\label{resf0} \\
F^{I=1}_{\pm}(x_{\rm Bj})&=&N_C\frac{M_N}{2\pi}
\langle N| D_{3i}
\int^\infty_{M_N|x_\mp|}p dp \int_0^{2\pi}d\Phi\
\nonumber \\* && \hspace{3cm}\times
\tilde\psi_{\rm v}^{\dag}(\mbox{\boldmath $p$}_\mp)\tau_i
\left(1\mp\alpha_3\right)\tilde\psi_{\rm v}(\mbox{\boldmath $p$}_\mp)
|N\rangle
{\rm tr}\left[\tau_3\left({{t_l t_m}\atop{t_m t_l}}\right)\right]\ ,
\label{resf1} 
\ee
where
$x_{\pm}=x_{\rm Bj}\pm\epsilon_{\rm v}/{M_N}$
and 
${\rm cos}(\Theta^\pm_p)={M_N}x_\pm/{p}$.
The polar--angle, $\Theta^\pm_p$, between $\mbox{\boldmath $p$}_\pm$ 
and $\mbox{\boldmath $q$}$ is fixed for a given value of the Bjorken 
parameter, $x_{\rm Bj}$. Hence the wave--function depends implicitly 
on $x_{\rm Bj}$ because 
$\tilde\psi_{\rm v}(\mbox{\boldmath $p$}_\pm)=
\tilde\psi_{\rm v}(p,\Theta^\pm_p,\Phi)$.

Turning to the evaluation of the nucleon matrix elements defined 
above we first note that the Fourier--transform of the wave--function 
is easily obtained because the angular parts are tensor spherical 
harmonics in both coordinate and momentum spaces. Hence, only the 
radial part requires numerical treatment. Performing straightforwardly
the azimuthal integrations in eqs (\ref{resf0}) and (\ref{resf1})
reveals that the isoscalar part, $F_1^{I=0}$, depends solely on the 
classical part of the valence quark wave--function, $\Psi_0$. Thus
$F_1^{I=0}$ is identical for all nucleon states. On the other hand
the isovector part, $F_1^{I=1}$, is linear in the angular velocity,
$\mbox{\boldmath $\Omega$}$. Since the $z$--direction is distinct 
the collective quantities appear as combinations of $D_{33}\Omega_3$ 
and $D_{3i}\Omega_i$. When quantizing the collective coordinates 
these combinations are substituted by the nucleon spin operator
yielding \cite{Ad83}
\be
\langle N|D_{33}\Omega_3|N\rangle = -\frac{I_3}{3\alpha^2}
\quad {\rm and} \quad
\langle N|D_{3i}\Omega_i|N\rangle = -\frac{I_3}{\alpha^2}\ .
\label{colmat}
\ee
Here $I_3=\pm(1/2)$ is the nucleon isospin projection and $\alpha^2$ 
refers to the moment of inertia of the soliton. For consistency we 
constrain it to the valence quark contribution, $\alpha^2_{\rm v}$,
{\it cf.} eq (\ref{adler1}). 
The isovector part is obviously proportional to the isospin projection 
but independent of the spin projection, as expected for unpolarized 
structure functions. It is convenient to define structure functions 
$f^{I}_{\pm}(x_{\rm Bj})$ with the nucleon matrix elements already 
computed via
\be
F^{0}_{\pm}(x_{\rm Bj}) = f^{0}_{\pm}(x_{\rm Bj})
\quad {\rm and} \quad
F^{1}_{\pm}(x_{\rm Bj}) = 2 I_3 f^{1}_{\pm}(x_{\rm Bj})\ .
\label{resff}
\ee

\bigskip
\leftline{\large\it 5. Results}
\smallskip

In figure 1 we display the unpolarized structure functions
$f^{0,1}_\pm$ for a constituent quark mass of $m=350{\rm MeV}$.
In that case the valence quark contribution to the moment 
of inertia is about 86\%. This shows that the vacuum is only 
moderately polarized and that the valence quark approximation is 
well justified. Here we assume the experimental value ($940{\rm MeV}$)
for the nucleon mass. We observe that the structure functions are 
well localized in the interval $0\le x_{\rm Bj}\le1$. The result
that the structure functions slightly exceed $x_{\rm Bj}=1$ is 
common to approaches which treat the nucleon as extended objects. In 
the context of bag type models various projection techniques have 
been proposed \cite{Ja80,Sc91,So94} to remedy this problem. This, 
however, is not the central issue of this paper.
\begin{figure}[t]
\centerline{
\epsfig{figure=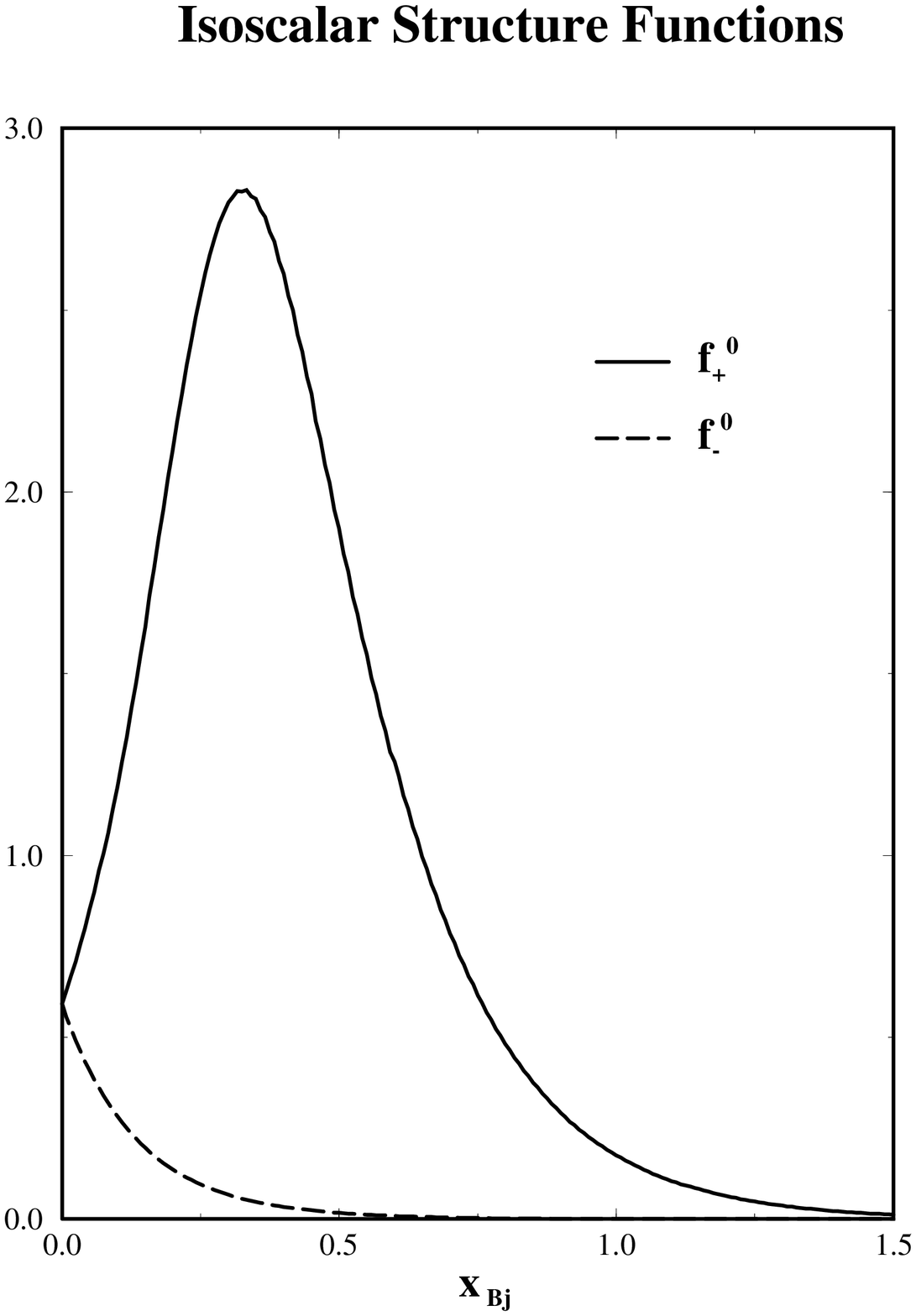,height=7.9cm,width=8.0cm}
\epsfig{figure=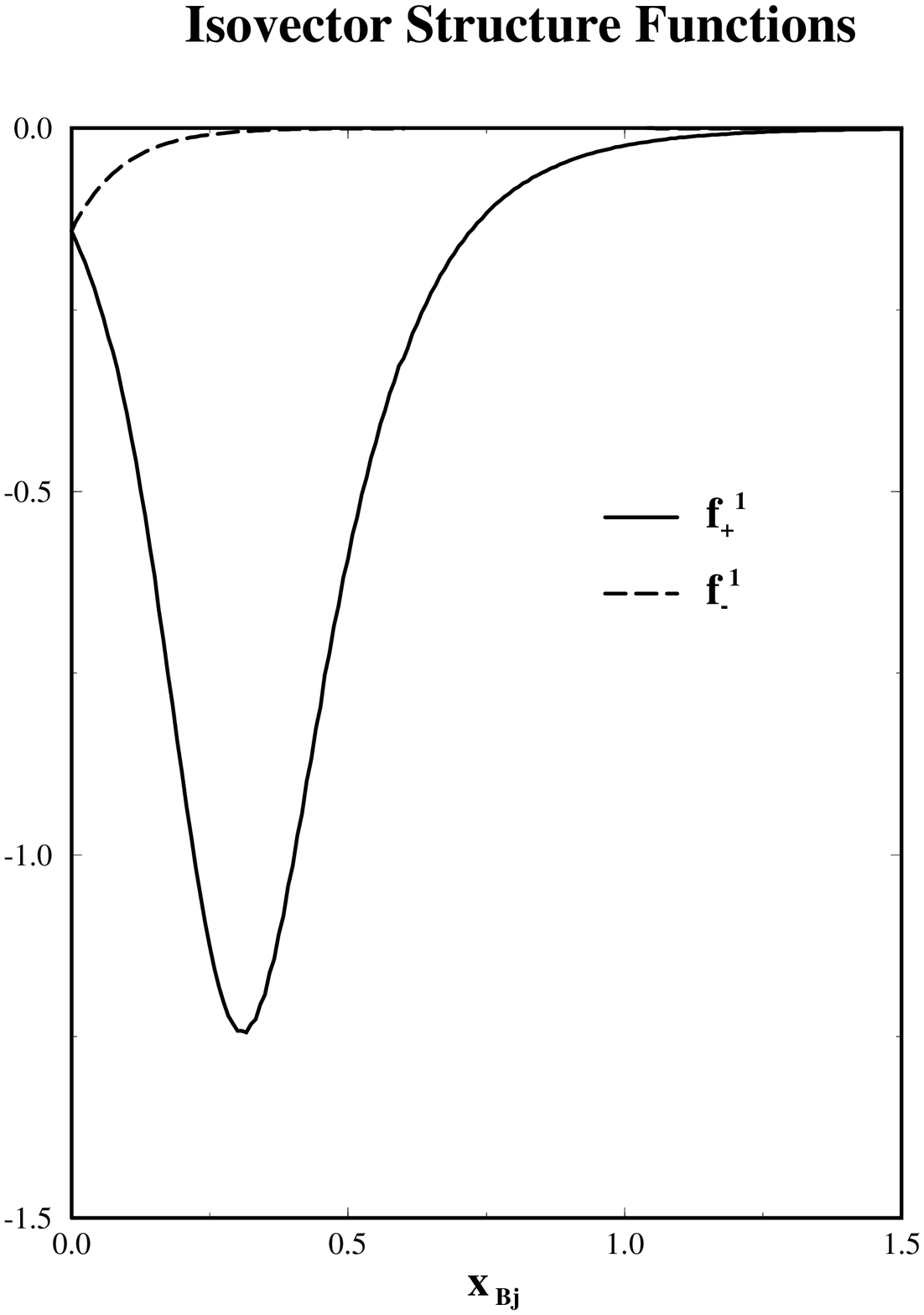,height=7.9cm,width=8.0cm}}
\vskip0.4cm
\caption{The unpolarized structure functions $f^{0,1}_\pm$ as 
functions of the Bjorken variable $x_{\rm Bj}$.}
\vskip-0.2cm
\end{figure}

The results displayed in figure 1 are the central issue of this 
calculation and it is of great interest to compare them with the 
available data. In this context we consider the structure functions 
for electron nucleon scattering. The associated isospin matrices 
are $t_a t_b=t_b t_a=(5+3\tau_3)/18$ yielding
\be
F_1^{eN}=\frac{1}{9}\left(5(f^0_+ - f^0_-) 
-6I_3(f^1_+ - f^1_-)\right)=\frac{1}{2x_{\rm Bj}}F_2^{eN}\ ,
\label{fen}
\ee
where the second equation results from the Callan--Gross 
relation. As for all effective low--energy models of the
nucleon, the predicted results are at a scale lower than
the experimental data. In order to carry out a sensible comparison
either the model results have to be evolved upward or the QCD
renormalization group equations have to be used to extract structure
functions at a low--renormalization point. The latter procedure 
has been employed in ref \cite{Gl95}\footnote{These authors also
provide a low scale parametrization of quark distribution functions.
However, these refer to perturbatively interacting partons. 
Distributions for the NJL--model constituent quarks could in 
principle be extracted from eqs. (\ref{resf0})--(\ref{resf1}). It 
is important to stress that these distributions may not be compared
to those of ref \cite{Gl95} because the associated quarks fields are 
different in nature.} to make available a low--scale parametrization
of the empirical data on $F_2^{eN}$. From figure 2 we observe that
the NJL model prediction for $F_2^{ep}-F_2^{en}$ reproduces the gross
features of this parametrization although the maximal value of the
prediction is a bit too large. On the other hand the low--scale value
are more enhanced at small $x_{\rm Bj}$. To illustrate the origin of 
the bumb at $x_{\rm Bj}\sim0$ we have also included the low--scale 
parametrization with the $\alpha_s$--corrections omitted, {\it cf.}
eq (7) of ref \cite{Gl95}. These are actually the starting point 
for computing the low--scale parametrization. When including the 
$\alpha_s$--corrections the integral (\ref{gottrule}) is forced to 
remain unchanged. As the $\alpha_2$--corrections shift the structure
functions to larger $x_{\rm Bj}$ the (artificial) bumb at 
$x_{\rm Bj}\sim0$ emerges. As an aside we would like to mention 
that the agreement between the NJL--model predictions and the 
parametrized structure functions is better when the
$\alpha_s$--corrections are omitted. This indicates that a 
fine--tuning of the low--scale momentum might improve the 
agreement even more.

With regard to the vacuum contribution it should be emphasized that
it will not simply add to the valence quark piece because when 
computing the isovector structure functions we have substituted 
$\alpha^2_{\rm v}<\alpha^2$ in eq (\ref{colmat}). For 
$m=400 (450){\rm MeV}$ we find $\alpha^2_{\rm v}/\alpha^2=78(72)\%$. 
We observe that for the combination $F_2^{ep}-F_2^{en}$ the 
agreement with the parametrization \cite{Gl95} improves as $m$ 
increases. We conjecture that this feature survives when the 
vacuum contribution is included because the moment of inertia 
enters the denominator of $F_2^{ep}-F_2^{en}$.  Furthermore the 
integral
\be
S_G=
\int_0^\infty \frac{dx_{\rm Bj}}{x_{\rm Bj}}
\left(F_2^{ep}-F_2^{en}\right) {\bjlim} \
2\int_0^\infty dx_{\rm Bj} \left(F_1^{ep}-F_1^{en}\right)=
0.29\  (0.27)
\label{gottrule}
\ee
agrees reasonably well with the empirical value $S_G=0.235\pm0.026$
\cite{Ar94} for the Gottfried sum rule. In particular the deviation 
from the na{\"\i}ve value (1/3) \cite{Go67} is in the correct 
direction. 
\begin{figure}[t]
\centerline{
\epsfig{figure=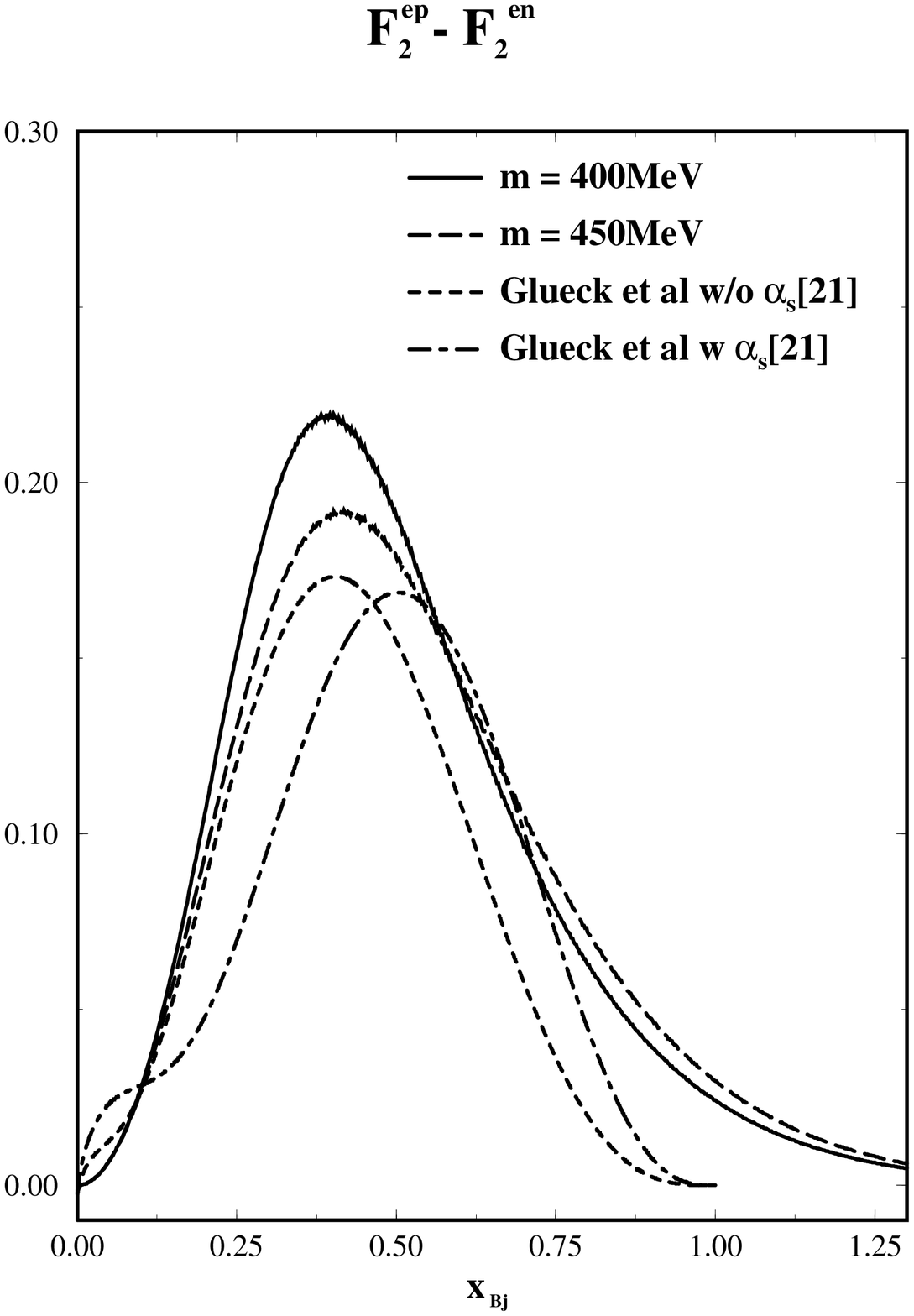,height=7.9cm,width=8.0cm}
\epsfig{figure=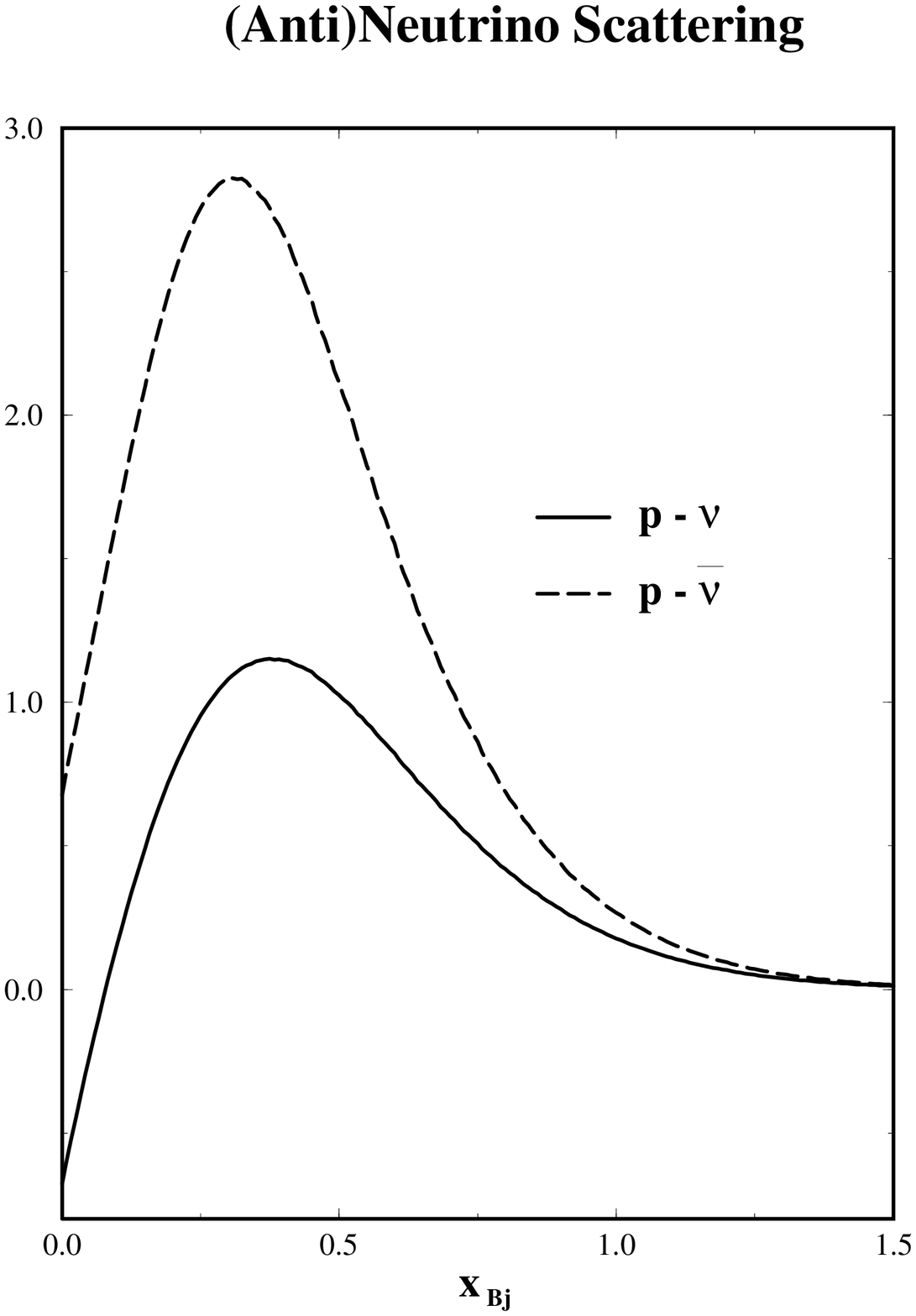,height=7.9cm,width=8.0cm}}
\vskip0.4cm
\caption{The valence quark approximation to the unpolarized 
structure functions as functions of the Bjorken variable 
$x_{\rm Bj}$. Left panel: The prediction on the Gottfried sum for 
two values of the constituent quark mass $m$. We compare with 
the low--scale parametrization of ref \protect\cite{Gl95}.
Right panel: $F_1^{{\nu}p}$ and $F_1^{{\bar \nu}p}$ for 
$m=450{\rm MeV}$.} 
\vskip-0.2cm
\end{figure}
 
For the weak scattering precesses ${\nu}p$ and ${\bar \nu}p$ we 
demand $t_a t_b=(1\pm\tau_3)/2$ yielding the linear combinations 
\be
F_1^{\nu p}=2\left(f_+^0-f_-^0+f_+^1+f_-^1\right)
\quad {\rm and} \quad
F_1^{{\bar \nu}p}=2\left(f_+^0-f_-^0-f_+^1-f_-^1\right)
\label{f1nup}
\ee
which are also plotted in figure 2. Although our 
wave--functions ({\it cf.} section 3) are quite different from those 
in the bag model the shape of the structure functions is similar. In 
particular the structure functions $F_1^{{\bar\nu}p,{\nu}p}$ do not 
vanish at $x_{\rm Bj}=0$ in both models. Despite that we essentially 
take only one quark eigenstate into account, 
we find a clear smearing of the structure functions. This shows that 
relativistic effects, {\it i.e.} a sizable lower component of the 
valence quark wave--function, play a significant role. These effects 
also cause the maximum of the structure to be shifted from 
$\epsilon_{\rm v}/M\approx0.26$ to about $0.37$. As in the bag 
model calculation of ref \cite{Ja75} we find that $F_1^{\nu p}$ is 
negative in the vicinity of $x_{\rm Bj}=0$. This appears to be linked 
to the omission of the vacuum states when computing the hadronic 
tensor (\ref{stpnt}).

Let us briefly comment on the Adler sum rule. Note that in the 
Bjorken limit, where the Callan--Gross relation is satisfied, the 
Alder and Bjorken sum rules are equivalent. It is an easy matter of 
exercise to verify that 
\be
\int_0^\infty dx \left[f_+^1(x)+f_-^1(x)\right] =
-\frac{N_C}{4\alpha^2}
\sum_\mu\frac{|\langle\mu|\tau_3|{\rm v}\rangle|^2}
{\epsilon_\mu-\epsilon_{\rm v}}
=-\frac{\alpha_{\rm v}^2}{2\alpha^2}\ .
\label{adler1}
\ee
Thus the Alder sum rule is satisfied once we assign the moment of 
inertia to its valence quark contribution, $\alpha^2_{\rm v}$. It is 
obvious that this sum rule will be recovered without this restriction 
when the contribution of the polarized vacuum is included in the 
evaluation of the functional trace (\ref{gendef1}).
The Adler sum rule also serves as a test for our numerical treatment. 
It furthermore manifests the parton model interpretation because
adopting unity as the upper boundary of the integral 
(\ref{adler1}) saturates this sum rule already by 99\% for the 
parameters used here. We should mention that the momentum sum rule 
is not satisfied in the valence quark approximation. The analytical 
proof of the momentum sum rule involves the classical equation of 
motion for the chiral field. As the polarized vacuum contributes 
to this equation it is obvious that including only the valence quark 
level in the calculation of the structure functions violates this 
sum rule. Numerically, however, this violation is small. For example,
for $m=450{\rm MeV}$ we are missing about 20\%. This number decreases 
with the constituent mass and may be interpreted as the momentum 
carried by the polarized vacuum\footnote{This consistency check 
requires one to use the soliton mass ($\sim 1.2{\rm GeV}$) rather 
than the experimental value for the nucleon mass.}.

\bigskip
\leftline{\large\it 6. Summary and Outlook}
\smallskip

The present study is intended as the first step towards clarifying 
the connection between the chiral soliton picture of the nucleon 
and the quark parton description. This has also to be regarded as 
an attempt to combine the phenomenologically successful concept of 
chiral symmetry with the quark parton model. For this purpose we have 
presented a first calculation of nucleon structure functions in the 
Bjorken limit from a chirally symmetric model. The mean field quark 
wave--functions in the background of a chiral soliton represent a 
non--trivial coupling of spin and isospin to the so--called grand 
spin. Baryon states possessing good spin and isospin are subsequently 
generated by cranking the soliton, see eq (\ref{valrot}). As a 
consequence there are rotational corrections to the mean field 
predictions of the structure functions. These corrections contribute 
to the isovector part of the unpolarized structure functions and 
are mandatory to reproduce the Adler sum rule. This form of the 
nucleon wave--function constitutes a major difference to quark 
models which are not based on a non--trivial chiral field as 
{\it e.g.} the bag model where baryons are described as direct 
products of spin and isospin eigenstates. In order to establish the 
connection between the two pictures we have (as a first step) 
restricted ourselves to the valence quark approximation. The results 
are in reasonable agreement with bag model results \cite{Ja75} (which 
does not include the Dirac sea either) and the empirical value of 
the Gottfried sum rule. Also the low--scale parametrization of the 
combination $F_2^{ep}-F_2^{en}$, which enters this sum rule,
is satisfactorily reproduced.

This encourages further studies in various directions. It is 
obvious that for the full computation of the structure functions
the polarization of the vacuum quark states has to be incorporated. 
Although for small constituent quark masses the soliton is dominated 
by the valence quark configuration it will be very interesting to
have available direct access to the sea quarks. The regularization
of the functional trace (\ref{gendef1}) will be rather involved. 
Fortunately the Adler and momentum sum rules may be employed to perform 
consistency checks. This study will illuminate whether (and how) the 
vacuum contribution violates the identity $f^I_+(0)=f^I_-(0)$, which 
is also observed in the bag model \cite{Ja75}. Within the parton model 
picture the sea quarks cause a violation of this relation.

Of special interest are the polarized structure functions which are 
to be extracted from the anti--symmetric part of the hadronic tensor 
$W^{(A)}_{\mu\nu}=(W_{\mu\nu}-W_{\nu\mu})/2$. The smallness of the 
first moment of the associated flavor singlet structure function is 
known as the proton spin puzzle. Since almost all chiral soliton models 
provide a reasonable explanation of this puzzle the computation of the 
entire structure function will provide further understanding how the 
nucleon is built up from its constituents.

One wonders whether the functional trace (\ref{gendef1}) has a suitable 
interpretation in chiral models with mesons as the fundamental fields.
Although it is possible to identify quark bilinear quantities in such 
soliton models via saturation of the Ward--identities, the analogues of 
the quark bilinears are always local. Hence models with 
{\it fundamental} meson fields contributing to the currents seem to be 
less tractable for calculating structure functions. In this respect an 
expansion of the functional trace (\ref{gendef1}) in derivatives of the 
chiral field might provide an effective operator suitable to compute 
structure functions in purely mesonic soliton models. However, first
investigations along this line have given disappointing results
\cite{Ja96}. This might be related to the failure of the gradient 
expansion in the soliton sector.

Gluonic effects are known to significantly contribute to the structure
functions, they may even cause some of them to be singular\footnote{For
example, the twist three spin average structure function $e(x_{\rm Bj})$
receives a pomeron contribution which behaves like 
$x_{\rm Bj}^{-2}$ \cite{Ja92}.} at $x_{\rm Bj}=0$. Such singularities 
will not appear in the soliton model calculation (neither do they in 
the bag model calculation \cite{Ja92}). Hence a further study of the 
structure functions may provide some insight how to effectively 
incorporate gluonic degrees of freedom in NJL--type models.

\bigskip
\leftline{\large\it Acknowledgements}
\smallskip

One of us (HR) gratefully acknowledges a fruitful discussion with
R. L. Jaffe. Further LG is grateful for insightful comments 
by G. R. Goldstein.

\end{document}